# Electronic structure, structural and optical properties of thermally evaporated CdTe thin films

S.Lalitha[1], S.Zh.Karazhanov[2,3], P.Ravindran[2*], S.Senthilarasu[1], R.Sathyamoorthy[1] and J.Janabergenov[4]

[1]R & D Department of physics, Kongunadu Arts and Science college, Coimbatore-641 029, Tamilnadu, India.
[2]Department of Chemistry, University of Oslo, PO Box 1033 Blindern, N-0315 Oslo, Norway.
[3]Physical-Technical Institute, 2B Mavlyanov St., 700084 Tashkent, Uzbekistan.
[4]Karakalpak State Pedagogical Institute, 104 Dosnazarov St.742005 Nukus, Karakalpakstan, Uzbekistan.
*Corresponding author: Fax: +47 22 85 54 41/55 65, E-mail:ponniah.ravindran@kjemi.uio.no

**Abstract**

Thin films of CdTe were deposited on glass substrates by thermal evaporation. From the XRD measurements it is found that the films are of zinc-blende-type structure. The lattice parameter was determined as $a = 6.529$ Å, which is larger than 6.48 Å of the powder sample, because the recrystallized lattice of the grown films is subjected to a compressive stress aroused as a result of the lattice mismatch and/or differences in thermal expansion co-efficient between the CdTe and the underlying substrate. Transmittance, absorption, extinction, and refractive coefficients are measured. Electronic structure, band parameters and optical spectra of CdTe were calculated from *ab initio* studies within the LDA and LDA+*U* approximations. It is shown that LDA underestimates the band gap, energy levels of the Cd-4$d$ states, $s$-$d$ coupling and band dispersion. However, it calculates the spin-orbit coupling correctly. LDA+*U* did not increase much the band gap value, but it corrected the $s$-$d$ coupling by shifting the Cd-4$d$ levels towards the experimentally determined location and by splitting the LDA-derived single $s$ peak into two peaks, which originates from admixture of $s$ and $d$ states. It is shown that the $s$–$d$ coupling plays an important role in absorption and reflectivity constants. The calculated optical spectra fairly agree with experimental data. Independent of wave-vector scissors operator is found to be a good first approximation to shift rigidly the band gap of CdTe underestimated by LDA.

## 1. Introduction

Thin films of II-VI semiconductors are currently used in many semiconductor devices such as photo-electrochemical cells, field effect transistors, detectors, photodiodes, photoconductors and photovoltaic solar cells [1-8]. CdTe has a direct band gap of 1.5 eV at room temperature and hence it is a suitable material for photovoltaic applications [9-13].

For fabrication of the CdTe films a variety of preparation techniques have been employed such as vacuum deposition [14-16], electro deposition [17], molecular beam epitaxy [18], metal-organic chemical vapor deposition [19], closed-space sublimation [20], and screen-printing [21]. The vacuum evaporation method has some advantages such as: the amount of impurities included in the growing layer will be minimized, the tendency to form oxides will be considerably reduced and finally straight line propagation will occur from the source to substrate. This method has been employed by us for preparation of the CdTe thin films. Because of the recrystallization, the lattice parameter of the CdTe sample shall be different from the commonly grown ones. Consequently, one can expect the band parameters to differ from the commonly used ones.

An extensive experimental investigation has been carried out on CdTe thin films, but a detailed knowledge about the effective masses at the main band extremum is required to understand the electronic structure of CdTe for its suitability in device applications. Although simple and formal expressions for the effective masses can readily be obtained from second order perturbation theory, the important parameters in the model have not been known with sufficient accuracy to render the validity of the theoretical results. The reason for this is depending on different approaches to the band structure calculations, mainly in the empirical methods, where the main objective is to reproduce energy values at high symmetry points where the correct dispersion of the bands is not well predicted.

The basic optical properties of semiconductors result from the electronic excitation in crystals when an electromagnetic wave is incident on them. As the optical spectra involves both valence band (VB) and conduction band (CB), the optical absorption measurements The theoretical and experimental investigations on the optical behavior of thin films deal primarily with optical reflection, transmission and absorption properties, and with their relation to the optical constants of the films. Moreover, the reflection, transmission, and interferometric properties of thin films have made it possible to determine the optical constants conveniently. The absorption studies have led to a variety of interesting thin film optical phenomena, which have thrown considerable light on the electronic structure of solids [22]. Absorption studies, on the other hand, provide a simple means of the evaluation of the band gap, optical transitions that may be direct or indirect, allowed or forbidden and also of the nature of the solid material [23]. CdTe possesses good absorption characteristics in the II-IV compound series.

This paper presents experimental and theoretical studies aimed at understanding of crystal structure, electronic properties, band parameters and optical properties of thermally evaporated CdTe thin films.

## 2. Methods
### 2.1. *Ab initio* calculations

The electronic band structure of CdTe in the zinc-blende structure is studied using the Vienna *ab initio* simulation package (VASP) [24] and full potential linear muffin-tin orbital (FP-LMTO) program [25], which calculate the Kohn-Sham eigenvalues within the framework of density functional theory (DFT) [26]. The calculations have been performed with the use of the local density approximation (LDA) [27] and the LDA plus multiorbital mean-field Hubbard potential (LDA+*U*) [28-30], which includes the on-site Coulomb interaction in the LDA Hamiltonian. The exchange and correlation energy per electron have been described by the Perdew-Zunger parametrization [31] with the quantum Monte Carlo procedure of Ceperley-Alder [32]. The interaction between electrons and atomic cores is described by means of non-norm-conserving pseudopotentials implemented in the VASP package [24]. The pseudopotentials are generated in accordance to the projector-augmented-wave (PAW) method [33,34]. The use of the PAW pseudopotentials addresses the problem of the inadequate description of the wave functions in the core region common to other pseudopotential approaches [35]. Also it allows us to construct orthonormalized all-electron-like wave functions for the Cd-4$d$, -5$s$ and Te -5$s$ and -5$p$ valence electrons.

The values of the parameters *U* and *J* required for the LDA+*U* studies were found from band structure calculations for different values of U by adjusting the position of the Cd-4$d$ bands to the experimentally established location [36]. The thus obtained empirical values of *U* were used to explore further the electronic structure and band parameters within the LDA+*U* procedure. Self-consistent calculations were performed using a 10x10x10 mesh frame according to the Monkhorst-Pack scheme. The completely filled semicore-Cd-4$d$ states have been considered as valence states.

For band structure calculations we used the experimentally determined lattice parameter $a = 6.529$ Å obtained from the present XRD measurements. The value of '$a$' is considerably higher than the other experimental data $a$ = 6.486 Å [37], $a$ = 6.48 Å [38,52], and $a$ = 6.482 Å [39]. So, for comparison, the calculations were also done for other $a$ = 6.486 Å taken from the Ref. [37]. The unit-cell vectors of the zinc-blende-type structures are $\mathbf{a}$ = (0, 1/2, 1/2) $a$, $\mathbf{b}$ = (1/2, 0, 1/2) $a$, $\mathbf{c}$ = (1/2, 1/2, 0) $a$, where '$a$' is the cubic lattice constant. There are four CdTe formula units per unit cell specified by Cd at (0, 0, 0) and Te at (1/4, 1/4, 1/4).

Effective masses of electrons and holes are calculated by:

$$\frac{1}{m_c(\mathbf{k})} = \frac{1}{\hbar^2} \frac{\partial^2 E(\mathbf{k})}{\partial \mathbf{k}^2}\bigg|_{\mathbf{k}=\mathbf{k}_0} \quad (1)$$

for a direction $\mathbf{k}$ about an extremum point $\mathbf{k}_0$ in the Brillouin zone. We have studied effective masses along [001], [011] and [111] directions in the vicinity of the $\Gamma$ point. The band energies have been calculated for a sequence of $\mathbf{k}$ points around $\Gamma$ and from that $m_c$ values have been calculated from Eq. (1) directly by polynomial fitting. However, as it was demonstrated for a number of III-V and II-VI semiconductors, the LDA calculations underestimate not only the band gap, but also the CB effective masses [40-45]. A simple way of correcting the LDA derived CB masses is to assume that LDA calculates the momentum matrix elements correctly [40,42,43,45]. This assumption was used in this work.

We have also considered the CB effective masses $m_c^{001}$, $m_c^{011}$, $m_c^{111}$, heavy-hole effective masses $m_{hh}^{001}$, $m_{hh}^{011}$, and $m_{hh}^{111}$, and light-hole effective masses $m_{lh}^{001}$, $m_{lh}^{011}$, and $m_{lh}^{111}$ for the above calculations. Throughout the paper the masses are presented in the unit of the free-electron rest mass $m_0$.

Imaginary part of the dielectric function $\varepsilon_2(\omega)$ was calculated by the DFT within LDA and LDA+$U$ approximations.

$$\varepsilon_2^{ij}(\omega) = \frac{Ve^2\mathbf{k}}{2\pi\hbar m^2\omega^2} \int d^3\mathbf{k} \sum_{nn'} \langle \mathbf{k}n|p_i|\mathbf{k}n^*\rangle\langle \mathbf{k}n^*|p_j|\mathbf{k}n\rangle \times \quad (2)$$
$$\times f_{\mathbf{k}n}(1-f_{\mathbf{k}n^*})\delta(f_{\mathbf{k}n^*} - f_{\mathbf{k}n} - \hbar\omega)$$

Here $(p_x, p_y, p_y) = p$ is the momentum operator, $f_{\mathbf{k}n}$ is the Fermi distribution, and $|\mathbf{k}n\rangle$ is the crystal wave function, corresponding to energy $\varepsilon_{\mathbf{k}n}$ with momentum $\mathbf{k}$. Due to the cubic structure of CdTe, the optical spectra are isotropic; consequently, only one component of the dielectric function is analyzed. The real part of the dielectric function $\varepsilon_1(\omega)$ is then calculated using the Kramers-Kronig transformation. These two spectra were then used to calculate all the other optical spectra. In this paper, we presented the absorption coefficient $\alpha(\omega)$, the reflectivity $R(\omega)$, as well as the refractive index $n(\omega)$ and the extinction coefficient $k_e(\omega)$ calculated using the following expressions:

$$R(\omega) = \left|\frac{\sqrt{\varepsilon(\omega)}-1}{\sqrt{\varepsilon(\omega)}+1}\right|^2, \quad (3)$$

$$\alpha(\omega) = \sqrt{2}\omega\sqrt{\sqrt{\varepsilon_1^2(\omega)+\varepsilon_2^2(\omega)} - \varepsilon_1(\omega)}, \quad (4)$$

$$n(\omega) = \sqrt{\frac{\sqrt{\varepsilon_1^2(\omega)+\varepsilon_2^2(\omega)} + \varepsilon_1(\omega)}{2}}, \quad (5)$$

$$k_e(\omega) = \sqrt{\frac{\sqrt{\varepsilon_1^2(\omega)+\varepsilon_2^2(\omega)} - \varepsilon_1(\omega)}{2}}. \quad (6)$$

Here $\varepsilon(\omega) = \varepsilon_1(\omega) + i\varepsilon_2(\omega)$ is the complex dielectric function. The optical spectra are calculated by DFT within LDA, GGA, and LDA+$U$ approximations for the energy range 0-20 eV.

It is well known that the optical spectra coming out from the calculations have many fine features, if they are not broadened [46-49]. To reproduce the experimental conditions more correctly, the calculated optical spectra are broadened. Lifetime broadening has been performed by convoluting the absorptive part of the dielectric function with a Lorentzian with 0.0 $(\hbar\omega)^2$ eV with quadratic scheme. Then the scissors operator was applied to adjust to the peaks on the optical spectra. It should also be noted that some smearing is caused by the finite instrumental resolution.

### 2.2. Structural and optical properties

Cadmium Telluride (Sigma-Aldrich 99.99+% purity) thin films were deposited on well cleaned glass substrates by thermal evaporation under a pressure of ~$10^{-6}$mbar. A quartz-crystal oscillation monitor determined the thickness of the film. The structure of the film was analyzed by using a X-ray diffractometer (JEOL-Japan, JDX8030 model) applying CuK$\alpha$ radiation.

The lattice parameter '$a$' has been evaluated from the relation

$$a = d\sqrt{h^2 + k^2 + l^2}, \quad (7)$$

where '$d$' is the atomic spacing value and $h$, $k$, and $l$ are the Miller indices.

For experimental studies of the optical spectra, the CdTe thin film was deposited onto a well-cleaned glass substrate by thermal evaporation. Optical transmittance and absorption measurements have been performed for the CdTe films. The results were recorded using a double beam JASCO V-570 Spectrophotometer in the wavelength range from 300 to 2500 nm. The absorption coefficient $(\alpha)$ was calculated from the transmittance spectra using the relation [12]

$$\alpha = \frac{4\pi k_f}{\lambda}. \quad (8)$$

The extinction coefficient $k_f$ coming out from experimental measurements was then calculated from the relation

$$k_f = \frac{2.303 \times \lambda \times \log(1/T_0)}{4\pi t}, \quad (9)$$

where $T_0$ is the transmittance, $t$ is the film thickness, and $\lambda$ is the wavelength of the incident radiation. The plot of the transmission $T$ against wavelength $\lambda$ is found to vary as [50]

$$T = \frac{16n_a n_g n^2 \exp(-\alpha t)}{R_1^2 + R_2^2 \exp(-2\alpha t) + 2R_1R_2 \exp(-\alpha t)\cos(4\pi nt/\lambda)}, \quad (10)$$

$$R_1 = (n+n_a)(n+n_g), \quad (11)$$

$$R_1 = (n-n_a)(n_g - n). \quad (12)$$

Here $n_a$, $n_g$, and $n$ are the refractive indexes of the film, air and substrate, respectively. Iterations were carried out until the desired convergence was achieved.

### 3. Results And Discussion
### 3.1. Structural Analysis

The X-ray diffractogram of the CdTe film of thickness 5000 Å is shown in Figure 1. XRD pattern exhibits polycrystalline nature and a major diffraction peak is observed at $2\theta$ =23.6° which corresponds to the cubic (111) orientation. The presence of the predominant peak at $2\theta$ =23.6° suggests that the CdTe film is of zinc blende structure with a preferential orientation along the (111) plane. No diffraction peak corresponding to metallic Cd, Te, or other compounds was observed. The (111) direction is the close packing direction of the zinc blende structure and this type of textured growth has often been observed in polycrystalline CdTe films grown on amorphous substrates [51].

The value of the lattice parameter $a$ = 6.529 Å is greater than that found for a powder sample 6.48 Å [38,52]. The higher value of '$a$' is due to the recrystallized lattice in the grown films. For as grown samples a large value of '$a$' was noticed which suggests that the film is subjected to a compressive stress in plane parallel to the surface of the substrate. This stress is caused by the lattice mismatch and/or differences in thermal expansion co-efficient between the CdTe and the underlying substrate. This result has influenced the band structure calculations and the dispersion in the effective masses. The lattice parameter $a$ = 6.529 Å obtained from our XRD measurement is used for the calculation of the band structure using DFT.

### 3.2 Measurements of optical spectra

Figure 2 shows the experimentally measured transmittance and absorption coefficient spectrum of the CdTe thin films of thickness 5000 Å. The transmittance decreases, while the absorption coefficient increases with increasing photon energy. The transmittance values in the visible region are very low and it tells about the nature of the film in the visible region. Interference fringes have been formed in the higher wavelength regions. It shows the clear tendency of the semiconducting nature of the CdTe thin films. The influence of the increasing free carrier concentration on the spectral dependence of transmittance in the near IR region manifests in a decrease of the transmittance due to the free carrier absorption [53]. High transmittance in a higher wavelength region and a sharp absorption edge were observed in the films.

From our earlier results [12,54], we have concluded that the CdTe thin film is exhibiting a direct transition. Figure 3 shows the plot between $(\alpha h \nu)^2$ versus $h\nu$ for a film of thickness 5000 Å. From the plot the band gap energy is calculated, which is equal to 1.534 eV. Figure 4 shows the measured extinction coefficient and refractive index as a function of the photon energy. The refractive index is very low at the fundamental band gap region. Such a low refractive index has been reported by Manifacier et. al. [55] for $SnO_2$ films. The variation may be due to the surface effect and volume imperfection on the microscale [56-60]. The extinction coefficient decreases with increasing incident photon wavelength. The refractive index and extinction coefficients are oscillatory in nature. The increase in the refractive index with wavelength is attributed to an increase in the crystallinity of the deposited films.

### 3.3. Band structure

Band structure calculations have been performed for CdTe within LDA and LDA+U approaches. Band gaps calculated including the spin-orbit (SO) coupling ($E_g^{SO}$) and neglecting it ($E_g$) as well as the SO coupling energy ($\Delta_{SO}$) are listed in Table I. The band gap calculated by LDA is severely underestimated in agreement with previous DFT calculations [44,61]. This underestimation is because of the well-known DFT eigenvalue problem. The $E_g$ value calculated using LDA+U method is slightly increased. The band gaps calculated by both LDA and LDA+U including the SO coupling into the calculations are smaller than those without SO coupling. The only parameter which agrees with experimental data is the SO splitting energy $\Delta_{SO}$. The experimentally determined value of $\Delta_{SO}$ is smaller than the one calculated from VASP by about 7.7 % for LDA and 9.4 % for LDA+U. The calculated value of $\Delta_{SO}$ using the FP-LMTO method deviates from experiment more than that from VASP. The values of $E_g^{SO}$ and $\Delta_{SO}$ calculated by VASP for the lattice parameter $a$ = 6.486 Å of Ref. [37] are larger by about 70 meV and 4 meV, respectively, compared to those for $a$=6.529 Å determined in this work from XRD measurements. So, the severe underestimation of $E_g^{SO}$ is not related to the lattice parameter value.

Band dispersions are presented in Fig. 5. Energy levels of the Cd-4d electrons calculated from pure LDA are found to be well below the topmost VB states. However, the Te-5s-peak is single and is located much below the energy levels of the Cd-4d electrons in agreement with LDA calculations [61]. However, this result contradicts experimental findings [36], showing two peaks at about $E_v$-10.8 eV and $E_v$-11.5 eV (Ref. [36]) separated by the Cd-4d band, and a low intensity shoulder in the range from about $E_v$-9.5 eV to $E_v$-10.0 eV Ref. [36,62]. To solve the discrepancy between LDA calculations and the experimental findings regarding the Cd-4d levels, LDA+SIC approach was used in Ref. [61], which allows adjusting the filled semicore Cd-4d levels to the experimentally determined position. In order to treat the position of Cd-4d electrons in the correct position we have used the LDA+U approach.

Band structure calculations have been performed for different values of U. It is found that good fitting of the Cd-4d bands to photoemission measurements in Ref. [36] will be achieved for $U \approx 7$ eV, if the SO coupling is included into the calculations. However, if the latter is neglected, then the empirical value of $U$ becomes much smaller ($U \approx 5$ eV). Further band structure studies within the LDA+U approach were done for the above empirical values of the parameter U. The Cd-4d levels were shifted toward lower energies and split the LDA-derived single s-peak into two, thus recovering the theoretical result of Ref. [61] by LDA+SIC approach and providing good agreement with experimental findings of Ref. [36,62]. This result indicates the presence of strong s-d coupling [61].

Without SO coupling the VB spectrum near the $\Gamma$ point originates from the six fold degenerate $\Gamma_{15}$ states. The SO coupling splits the $\Gamma_{15}$ level into a fourfold degenerate $\Gamma_8$ (hh and lh) and doubly degenerate $\Gamma_7$ (sh) levels located in the order of decreasing energy ($\Gamma_8 > \Gamma_7$). Fine structure of the topmost VB in the close vicinity of the $\Gamma$ point is studied by both LDA and LDA+U methods with and without the SO coupling. The results are presented in Fig. 6. It is seen that there is almost no change in the LDA-and LDA+U derived band dispersions in the close vicinity of the $\Gamma$-point. This result confirms once again that the p-d coupling is not strong enough to change drastically the VB dispersion and carriers effective mass values.

Figure 7 presents the site and orbital projected density of states (DOS) for CdTe calculated within the LDA and LDA+U. Analysis shows that in the LDA calculations VB consists of four bands: the lowest one originates from Te-5s states slightly hybridised by Cd-4d ones. The next higher energy band basically is contributed from Cd-4d bands. It is slightly hybridised with Te-5s and -5p bands. It consists of well-defined two peaks: lower energy one is very sharp and intensive, while magnitude of the other peak located at higher energy is smaller. The next smaller peak is located at higher energy and is originated from hybridised Cd-5s- and Te-5p states. The topmost VB and lowest CB are not hybridised. The former is contributed from the Te-5p–states, while the latter comes from Cd- and Te-5s states.

In the LDA+U calculations the lowest VB and CB are changed drastically, while the other bands did not change much. The LDA-derived lowest two VB have been split into well-defined separated each from other bands. Compared to those calculated within LDA, these bands are strongly hybridised and all of them consisted of admixture of the Te-5s, -5p, and CdTe-4d states. LDA+U almost did not change contribution of the Te-5p states, while the s–d-s hybridisation becomes much more stronger. The two small intensity peaks located at the lower and higher energy sides of the central very sharp and intensive peak are almost equally contributed from the Te-5s and Cd-4d states. Locations of the lowest VB peaks coming out from present LDA and LDA+U calculations agree well with those of Ref. [61].

### 3.4. Effective masses

The CB effective masses $m_c^{100}$, $m_c^{110}$, $m_c^{111}$, heavy-hole effective masses $m_{hh}^{100}$, $m_{hh}^{110}$, $m_{hh}^{111}$, light-hole effective masses $m_{lh}^{100}$, $m_{lh}^{110}$, $m_{lh}^{111}$, and split-off hole effective masses $m_{sh}^{100}$, $m_{sh}^{110}$, $m_{sh}^{111}$ are calculated by DFT within LDA with and/or without the SO couplings (Tables II). Analysis shows that the effective masses along all the directions are more or less isotropic, which is related to symmetry properties of the crystal structure. Inclusion of the SO coupling into DFT calculations did not change much the effective masses. The reason is because dispersion corresponding to the CB, heavy- and light-hole bands in close vicinity of the $\Gamma$ point calculated with and without SO coupling [Fig. 5 (b) and Fig. 6] is almost the same.

The effective masses were not calculated for band structures derived from LDA+U studies, because the Cd-4d levels are located much below the topmost VB (Fig. 6) and, respectively, the p-d coupling is expected to be small enough to change the dispersion and VB masses. However, the CB masses are expected to be increased in the LDA+U calculations compared to the LDA derived ones. The CB electron effective masses are more than 3 times underestimated with respect to the experimentally determined one. This result is in contrast to that of Ref. [44], which overestimated the CB mass in the DFT calculations. The heavy-hole masses are <2 times smaller than the experimental data. However, light-hole effective masses are about < 3 times overestimated. Analysis of literature (see, e.g., Refs.[40,44] and references therein) shows that the effective masses calculated by different methods and softwares often differ each from other.

To fix the well-known disagreement between the calculated CB effective masses with experimental data the k.p theory [42,43,63,64] is used. The theory provides the most efficient and accurate expression to calculate the effective mass

$$\frac{1}{m_c} = 1 + 2\sum_i \frac{P_i^2}{E_c - E_i} = 1 + \frac{E_p}{E_g} + C, \quad (13)$$

$$C \equiv \sum_{i > \Gamma_{1c}} \frac{E_p^i}{E_g^i}. \quad (14)$$

Here $m_c$ is the lowest CB mass. Since the CB minimum is isotropic, $m_c = m_c^{100} = m_c^{110} = m_c^{111}$. $E_c - E_i$ is the energy gap between the $i^{th}$ CB state and the lowest one, and $P_i$ is the momentum matrix element between these states. $E_g$ is the fundamental direct $p$-$s$ energy gap and $E_p$ is the Kane momentum matrix element (defined as $2P^2/m_0$ in eV). Note that all the four basic parameters in Eq. (13) can be directly calculated by DFT: $E_g$ from the eigenvalues, $E_p$ from the wave functions, $m_c$ from the dispersion of the eigenvalues, and $C$ from Eq. (14) by first finding the DFT $E_p^i$ and $E_g^i$. $C$ is the constant that accounts for the contribution from the remote CB.

It should be noted that Eq. (13) is derived on the assumption that the pseudopotential is local $\hat{V}_{loc}$. However, if a non-local part of the pseudopotential $\hat{V}_{nloc}$ is included into consideration, then the Hamiltonian $\hat{H}$ becomes of the form

$$\hat{H} = -\frac{\hbar^2 \nabla^2}{2m} + \hat{V}_{loc}(r) + \hat{V}_{nloc}(r). \quad (15)$$

Then the k.p theory equation [Eq. (13)] was modified by Rieger and Vogl [65] for empirical non-local, relativistic pseudopotential theory, which was generalized afterwards by Pickard and Payne [66] by including the Vanderbilt's non-local and norm-conserving ultrasoft pseudopotentials and plane waves. The use of the non-local pseudopotentials, while leading to efficient and accurate results, introduces some complications in comparison to a purely local Hamiltonian - namely, the Hamiltonian no longer commutes with the position operator.

$$[\hat{V}_{loc}, \hat{r}] = 0, \quad (16)$$

$$i[\hat{H}, \hat{r}] = i\left[-\frac{\hbar^2}{2m_0}\nabla^2, \hat{r}\right] \frac{\hbar}{m_0}\hat{p} + i[\hat{V}_{nloc}, \hat{r}], \quad (17)$$

$$[[\hat{H}, \hat{r}], \hat{r}] = \left[\left[-\frac{\hbar^2}{2m_0}\nabla^2, \hat{r}\right], \hat{r}\right] \frac{\hbar^2}{2m_0} + [[\hat{V}_{nloc}, \hat{r}], \hat{r}]. \quad (18)$$

Then, following the procedure of Ref. [66] using the second order k.p perturbation theory, one can get for the effective mass the following equation

$$\frac{1}{m_c} = 1 + \frac{m_0}{\hbar^2}\langle\phi_n^{(0)}|i[[\hat{V}_{nloc},\hat{r}],\hat{r}]|\phi_n^{(0)}\rangle + 2\sum_i \frac{P_i^2}{E_c - E_i}. \quad (19)$$

Eq. (19) contains the contribution from the non-local part of the pseudopotential. Rieger and Vogl [65] have found that the non-local part of the pseudopotential contributes up to 20% to the calculated CB masses. Comparing Eq. (13) with Eq. (19) one can see that these two equations are similar. So the parameter $C$ can be regarded as the constant that accounts for the contribution of not only the remote CB, but also of the non-local part of the pseudopotential. The value of the parameter $C$ is estimated empirically in Ref. [64] to be equal to $C \approx -2$. This empirical value was shown to provide good agreement of the CB effective masses calculated by Eq. (13) with those determined experimentally for a number of semiconductors (see e.g. Ref. [64]). In this work, this value is used together with Eq. (13).

It was assumed in Refs. [42,43] that within the DFT the momentum matrix elements $E_p$ in Eq. (13) are calculated correctly. If so, then in Eq. (13) one can replace the band gap $E_g$ calculated by DFT with experimentally determined one $E_g^{Expt}$ and find the normalized CB effective mass $m_c^*$

$$\frac{1}{m_c^*} = 1 + C + \frac{E_g}{E_g^{Expt}}\left(\frac{1}{m_c} - 1 - C\right). \quad (20)$$

The CB effective masses $m_c^*$, calculated by Eq. (20) using the values of the LDA derived $E_g$ and $m_c$, the empirically determined $C$, and the experimental band gap $E_g^{Expt}$ are equal to 0.224, if the SO coupling in included and are equal to 0.103, when the SO coupling is neglected. It is seen that the latter shows good agreement with experimentally determined CB mass $m_c^{Expt}$, while the former is two times large than $m_c^{Expt}$. The possible reasons leading to the large discrepancy can be related to invalidity of the assumption of correctness of the momentum matrix elements, calculated by DFT, to deviation of the contribution of the remote band and non-local effects from the empirical value $C \approx -2$, and to deficiency of DFT in description of not only the band gap, but also the band dispersion etc.

3.5. *Ab initio* studies of the optical properties of CdTe

The response functions $\varepsilon_1(\omega)$ and $\varepsilon_2(\omega)$, absorption coefficient $\alpha(\omega)$, reflectivity $R(\omega)$, refractive index $n(\omega)$, and extinction coefficient $k_e(\omega)$ calculated within LDA and LDA+$U$ approximations for CdTe are displayed in the figure 8. The results are compared with standard experimental data for CdTe [67]. Locations of all the peaks in the spectral distribution of the optical spectra calculated by DFT are shifted toward lower energies compared to experimentally measured ones. A rigid shift toward higher energies has been performed in order to correct the DFT underestimation of the band gaps. In the thus obtained optical spectra all peak locations fairly agree with the experimentally determined ones. Consequently, the k-independent scissors operator can be applied to correct the band gap underestimated by DFT.

Analysis of the spectral distribution of the optical spectra shows that magnitude of the peaks corresponding to the fundamental absorption gap is overestimated compared to the experimental data. It can be related to overestimation of the momentum matrix elements $E_p$ (Eq. (20)), neglect of the Coulomb interaction between free electrons and holes (excitons), local-field and finite lifetime effects. Furthermore, for calculations of the imaginary part of the dielectric response function optical transitions from occupied to unoccupied states with fixed k vector are considered only. Also, the experimental resolution smears out many fine features. The assumption, that overestimation of the peaks is related to that of $E_p$ supports our conclusion that the renormalization technique used in this work can not be applied for the CdTe films considered.

It is found that the location of the peaks in the absorption and reflectivity spectra calculated and measured in this work, and that of Ref. [67] deviate each from other. The reason of the discrepancy can be related to the difference in the lattice parameters of the CdTe thin films studied in this work and that in Ref. [67]. The refractive index is shown to be small at the fundamental band gap region, which can be related to the surface effect and volume imperfection. The extinction coefficient decreases with increasing the incident photon wavelength, while the refractive index and extinction coefficients oscillate. The increase of the refractive index with increasing the wavelength is attributed to an increase in the crystallinity of the deposited films.

It is seen that the optical spectra derived from LDA and LDA+$U$ are almost the same except some differences for the absorption coefficient $\alpha(\omega)$ and reflectivity $R(\omega)$ at higher energies. The optical spectra calculated within these two approximations are compared to each other (Figure 9). Indeed, at the energy range 8-16 eV the values and peaks for $\alpha(\omega)$ and $R(\omega)$ differ each from other. In the absorption and reflectivity coefficients (Figure 8) calculated within LDA, two peaks are seen at energies 13.44 and 14.65 eV. Comparing with the DOS (Figure 7), it is thought that these peaks are coming from the single Te-5s and single Cd-4d states. However, in the LDA+U calculations these two optical spectra show three peaks at energies 13.36, 13.93, and 14.54 eV. By comparing with Figure 7, one can ascribe the origin of the first and third peaks to the splitted s band, while that at 13.93 eV comes from the hybridized Cd-4d states. In the experiments of Ref. [67] these peaks are smeared and not seen.

Note, that in the above optical spectra SO coupling was not included. The optical spectra were then calculated by the FP-LMTO method by LDA and LDA+U approximations, taking the SO coupling into account, and neglecting it. The results are presented in Figure 10. Analysis shows that LDA+U calculations by the FP-LMTO method without SO coupling show better agreement with the experimental data than those calculated with the VASP package. LDA+U studies by FP-LMTO including the SO coupling into calculations deviate significantly from experimental data. The discrepancy is more pronounced than that in the calculations neglecting the SO coupling. Especially, this problem is well pronounced for $\alpha(\omega)$.

**4. Conclusion**

Cadmium Telluride (Sigma-Aldrich 99.99+% purity) thin films were deposited on well cleaned glass substrates by thermal evaporation under a pressure of ~ $10^{-6}$ mbar. By analysis of the X-ray diffractogram it is found that the CdTe film exhibits a polycrystalline nature. The predominant peak at $2\theta$=23.6° indicates that the CdTe film is of zinc-blende structure with a preferential orientation along the (111) plane. The measured lattice parameter is $a$ = 6.529 Å, which is larger than that of the powder sample 6.48 Å [52]. The large value of '$a$' is found to be related to the recrystallized lattice in the grown films. For as grown samples the large value of '$a$' was found to be due to a compressive stress in plane parallel to the surface of the substrate caused by the lattice mismatch and/or differences in thermal expansion co-efficient between the CdTe and the underlying substrate. Using the measured lattice parameter, band structure parameters for CdTe have been obtained using first-principles calculations with the LDA and LDA+U approximation. It is shown that the band gap, energy levels of the Cd-4d states, band dispersion calculated by LDA and LDA+U are severely underestimated in agreement with previous *ab initio* studies. The experimentally determined value of the SO splitting energy is found to be smaller than the calculated ones using VASP, by about 7.7 % for LDA and 9.4 % for LDA+U. The band parameters underestimated by LDA were not increased much by LDA+U calculations; because the Cd-4d levels are located much below the top most VB states and as a result the p-d coupling is not strong enough. However, the s-d coupling is found to be significant, which is underestimated within LDA calculations showing well separated s and d bands, in disagreement with the photoemission measurements. This discrepancy was removed by the LDA+U approximation, which splits the LDA-derived single s peak into two peaks separated by the Cd-4d levels in agreement with previous theoretical and experimental studies. The origin of the peaks is found to be admixture of s and d states. The calculated carrier effective masses were compared with the available experimental measurements. It is shown that the CB effective masses calculated by DFT are more than three times smaller than the experimentally determined one. Using the second-order k.p perturbation theory the CB masses are normalized based on the assumption that DFT calculated correctly the momentum matrix elements. The CB mass calculated by this way neglecting the SO coupling agrees well with the experimental mass, while that accounting for the SO coupling exceeds it more than two times. Optical properties of thin film CdTe were studied by DFT within LDA and LDA+U approximations with and without SO coupling. It is found that location of the peaks in the calculated optical spectra agree well with experimentally determined one, whereas deviations are found for the absorption and reflectivity constants from experimental data in the energy range 5-12 eV. By comparing the optical spectra calculated by LDA and LDA+U it is found that the latter shows three peaks corresponding to the hybridized s–d band, which are not in the LDA calculations underestimating the s–d coupling. FP-LMTO calculations within LDA+U neglecting the SO coupling is found to be in better agreement with experimental data than those accounting for the SO coupling.

**Acknowledgments**

This work has received financial and supercomputing support from the Research Council of Norway and of the Academy of Sciences of Uzbekistan (project N31-36). SZK thanks R. Vidya, P. Vajeestan, and A. Klaveness (Department of Chemistry, University of Oslo, Oslo, Norway) for assistance in computations. Part of the experimental work has been supported by the University Grants Commission (UGC), India by awarding UGC-Research Award (Project No. F-30-1/2004 (SA-II)) to one of the author R.Sathyamoorthy.

**References**

[1] H. Hernandez-Contreras, C. Mejia-Garcia, and G. Contreras-Puente, Thin Solid Films 451-452 (2004) 203.
[2] J. Fritsche, T. Schulmeyer, A. Thiben, A. Klein, and W. Jaegermann, This Solid Films 431-432 (2003) 267.
[3] A. Romeo, D. L. Batzner, H. Zogg, and A. N. Tiwari, Thin Solid Films 361-362 (2000) 420.
[4] R. K. Sharma, G. Singh, and A. C. Rastogi, Sol. Energy Mater. Sol. Cells 82 (2004) 201.
[5] S. Seto, S. Yamada, and K. Suzuki, Sol. Energy Mater. Sol. Cells 67 (2001) 167.
[6] M. Becerril, O. Zelaya-Angel, J. R. Vargas-García, R. Ramírez-Bon, and J. González-Hernández, J. Phys. Chem. Solids 62 (2001) 1081.
[7] M. Y. E. Azhari, M. Azizan, A. Bennouna, A. Outzourhit, E. L. Ameziane, and M. Brunel, Sol. Energy Mater. Sol. Cells 45 (1997)
[8] J. J. Loferski, J. Appl. Phys. 27 (1956) 777.
[9] M. B. Das, S. V. Krishnaswamy, R. Petkie, P. Swab, and K. Vedam, Solid-State Electron. 27 (1984) 329.
[10] J. J. Wysocki and P. Rappaport, J. Appl. Phys. 31 (1960) 571.
[11] R. H. Williams and M. H. Patterson, Appl. Phys. Lett. 40 (1982) 484.
[12] S. Lalitha, R. Sathyamoorthy, S. Senthilarasu, A. Subbarayan, and K. Natarajan, Sol. Energy Mater. Sol. Cells 82 (2004) 187.
[13] H. Uda, S. Ikegami, and H.Sonomura, Jpn. J. Appl. Phys. 29 (1990) 2003.
[14] U. Khairnar, D. Bhavsar, R. Vaidya, and G. Bhavsar, Mater. Chem. Phys. 80 (2003) 421.
[15] N. Bakr, J. Cryst. Growth 235 (2002) 217.
[16] R. Sathyamoorthy, S. Narayandass, and D. Mangalaraj, Sol. Energy Mater. Sol. Cells 76 (2003) 217.
[17] X. Mathew, N. Mathews, P. Sebastian, and C. Flores, Sol. Energy Mater. Sol. Cells 81 (2004) 397.
[18] S. Ringel, A. Smith, M. MacDougal, and A. Rohatgi, J. Appl. Phys. 70 (1991) 881.
[19] T. Chu, S. Chu, C. Ferekides, J. Britt, and C. Wu, J. Appl. Phys. 71 (1992) 3870.
[20] G. Hernández, X. Mathew, J. Enráquez, B. Morales, M. Lira, J. Toledo, A. Juárez, and J. Campos, J. Mater. Sci. 39 (2004) 1515.
[21] A. Nakano, S. Ikegami, H. Matsumoto, H. Uda, and Y. Komatsu, Solar cells 17 (1986) 233.
[22] K. L. Chopra, Thin Film Phenomena, McGraw-Hill, New York, 1969.
[23] A. Goswami, Thin Film Fundamentals, New Age International (P) Ltd, Publishers, 1996.
[24] G. Kresse and J. Furthmüller, Phys. Rev. B 54 (1996) 11169.
[25] S. Y. Savrasov, Phys. Rev. B 54 (1996) 16470.
[26] P. Hohenberg and W. Kohn, Phys. Rev. B 136 (1964) B864.
[27] W. Kohn and L. J. Sham, Phys. Rev. B 140 (1965) A1133.
[28] V. I. Anisimov, I. V. Solovyev, M. A. Korotin, M. T. Czyzyk, and G. A. Sawatzky, Phys. Rev. B 48 (1993) 16929.

[29] S. L. Dudarev, G. A. Botton, S. Y. Savrasov, C. J. Humphreys, and A. P. Sutton, Phys. Rev. B 57 (1998) 1505.
[30] O. Bengone, M. Alouani, B. B. ochl, and J. Hugel, Phys. Rev. B 62 (2000) 16392.
[31] J. P. Perdew and A. Zunger, Phys. Rev. B 23 (1981) 5048.
[32] D. M. Ceperley and B. J. Alder, Phys. Rev. Lett. 45 (1980) 566.
[33] P. E. Blöchl, Phys. Rev. B 50 (1994) 17953.
[34] G. Kresse and D. Joubert, Phys. Rev. B 59 (1999) 1758.
[35] B. Adolph, J. Furthmüller, and F. Bechstedt, Phys. Rev. B 63 (2001) 125108.
[36] T. Loher, Y. Tomm, C. Pettenkofer, A. Klein, and W. Jaegermann, Semicond. Sci. Technol. 15 (2000) 514.
[37] O. Madelung and M. Schulz, eds., Numerical Data and Functional Relationships in Science and Technology. New Series. Group III: Crystal and Solid State Physics. Semiconductors.Supplements and Extensions to Volume III/17. Intrinsic Properties of Group IV Elements and III-V, II-VI and I-VII Compounds, Vol. vol. 22a, Springer, Berlin, 1982.
[38] N. K. Abrikosov, V. B. Bankina, L. V. Poretskaya, L. E. Shelimova, and E. V. Skudnova, Semiconducting II-VI, IV-VI, and V-VI compounds, Plenum, New York, 1969.
[39] O. Madelung, ed., Data in Science and Technology. Semiconductors: Other than Group IV Elements and III-V Compounds Springer, Berlin, Springer, Berlin, 1992.
[40] K. Kim, W. R. L. Lambrecht, B. Segall, and M. v. Schilfgaarde, Phys. Rev. B 56 (1997) 7363.
[41] L. C. L. Y. Voon, M. Willatzen, M. Cardona, and N. E. Christensen, Phys. Rev. B 53 (1996) 10703.
[42] M. Willatzen, M. Cardona, and N. E. Christensen, Phys. Rev. B 51 (1995) 13150.
[43] M. Willatzen, M. Cardona, and N. E. Christensen, Phys. Rev. B 51 (1995) 17992.
[44] S. Z. Karazhanov and L. C. L. Y. Voon, Semicond 39 (2005) 177.
[45] M. Oshikiri, F. Aryasetiawan, Y. Imanaka, and G. Kido, Phys. Rev. B 66 (2002) 125204.
[46] P. Ravindran, A. Delin, R. Ahuja, B. Johansson, S. Auluck, J. Wills, and O. Eriksson, Phys. Rev. B 56 (1997) 6851.
[47] P. Ravindran, A. Delin, P. James, B. Johansson, J. Wills, R. Ahuja, and O. Eriksson, Phys. Rev. B 59 (1999) 15680.
[48] P. Ravindran, A. Delin, B. Johansson, O. Eriksson, and J. Wills, Phys. Rev. B 59 (1999) 1176.
[49] A. Delin, P. Ravindran, O. Eriksson, and J. Wills, Int. J. Quantum Chem. 69 (1998) 349.
[50] J. C. Manifacier, J. Gasiot, and J. P. Fillard, J. Phys.: E: Sci. Instrum. 9 (1976) 1002.
[51] G. Gordillo, J. M. Flrez, and L. C. Hernández, Sol. Energy Mater. Sol. Cells 37 (1995) 273.
[52] M. A. Redwan, E. H. Aly, L. I. Soliman, A. A. El-Shazely, and H. A. Zayed, Vacuum 69 (2003) 545.
[53] J. Touṣkova, J. Kovanda, L. Dobiaṣova, V. Paṛizek, and P. Kielar, Sol. Energy Mater. Sol.Cells 37 (1995) 357.
[54] R. Sathyamoorthy, S. K. Narayandass, and D. Mangalaraj, Sol. Energy Mater. Sol. Cells 76 (2003) 339.
[55] J. C. Manifacier, M. D. Murcia, J. P. Fillard, and E. Vicario, Thin Solid Films 41 (1977) 127.
[56] J. M. Pawlikowski, Thin Solid Films 127 (1985) 29.
[57] E. Khawaja and S. G. Tomlin, J. Phys. D: Appl. Phys. 8 (1975) 581.
[58] P. Rouard and A. Meessen, Progress in Optics 15 (1977) 77.
[59] I. Filinski, Phys. Status Solidi B 49 (1972) 577.
[60] D. T. F. Marple, J. Appl. Phys. 35 (1964) 539.
[61] C. Persson and A. Zunger, Phys. Rev. B 68 (2003) 073205.
[62] A. K. J. Fritsche, A. Thißen, and W. Jaegermann, Thin Solid Films 387 (2001) 158.
[63] P. Lawaetz, Phys. Rev. B 4 (1971) 3460.
[64] C. Hermann and C. Weisbuch, Phys. Rev. B 15 (1977) 823.
[65] M. M. Rieger and P. Vogl, Phys. Rev. B 48 (1993) 14276.
[66] C. J. Pickard and M. C. Payne, Phys. Rev. B 62 (2000) 4383.
[67] S. Adachi, ed., Optical Constants of Crystalline and Amorphous Semiconductors. Numerical Data and Graphical Information, Kluwer Academic Publishers, Boston/Dordrecht/London, 1999.

| Method | $E_g$ | $E_g^{SO}$ | $\Delta_{SO}$ |
|---|---|---|---|
| LDA (VASP) | 0.4832 | 0.2055 | 0.8613 |
| LDA+U (VASP) | 0.6844 | 0.4044 | 0.8752 |
| LDA (FP-LMTO) | | 0.1571 | 1.0051 |
| LDA+U (FP-LMTO) | 0.9310 | 0.3674 | 1.0285 |
| Theory, Ref.61 | 0.4200 | | |
| Expt., Ref. 39 | | 1.6060 | 0.8000 |

**Table 1:** Band gaps [ $E_g$ and $E_g^{SO}$ ] and the SO splitting energy $\Delta_{SO}$ calculated within VASP [24] and FP-LMTO [25, 29] packages. The results are compared to the theory of Ref. [44] and experiment of Ref. [39].

| | $m_e^{100}$ | $m_e^{110}$ | $m_e^{111}$ | $m_{hh}^{100}$ | $m_{hh}^{110}$ | $m_{hh}^{111}$ | $m_{lh}^{100}$ | $m_{lh}^{110}$ | $m_{lh}^{111}$ |
|---|---|---|---|---|---|---|---|---|---|
| Without SO coupling | 0.03 | 0.03 | 0.03 | 0.39 | 0.46 | 0.54 | 0.33 | 0.31 | 0.35 |
| | 0.12 | 0.12 | 0.12 | 1.07 | 1.07 | 2.57 | 0.13 | 0.12 | 0.11 |
| Theory, Ref. 44 | 0.02 | 0.02 | 0.03 | 0.40 | 0.41 | 0.53 | 0.31 | 0.30 | 0.31 |
| With SO coupling | 0.09 | 0.10 | 0.10 | 0.72 | 0.81 | 0.84 | 0.13 | 0.12 | 0.12 |
| Expt. 39 | | | | | | | | | |

**Table 2:** Effective masses of electrons and holes (in units of the free-electron mass $m_0$) for CdTe calculated from band structure studies by VASP. The results are compared to experimentally determined data (Ref. [39]) and theoretical calculations of Ref. [44] by plane-wave pseudopotential method neglecting the SO couplings.

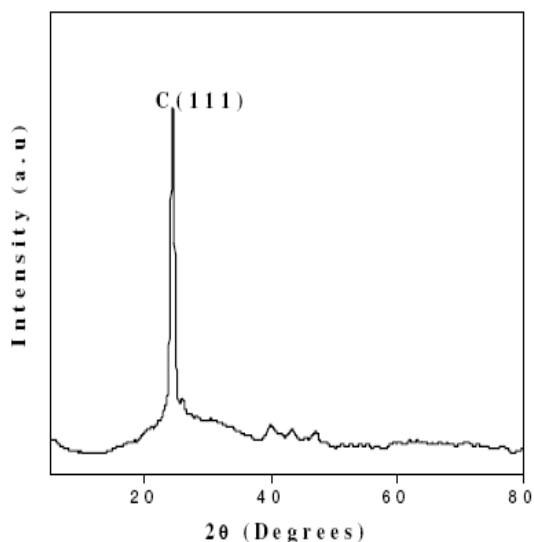

**Figure 1:** X-ray diffraction spectrum of CdTe thin film of thickness 5000 Å.

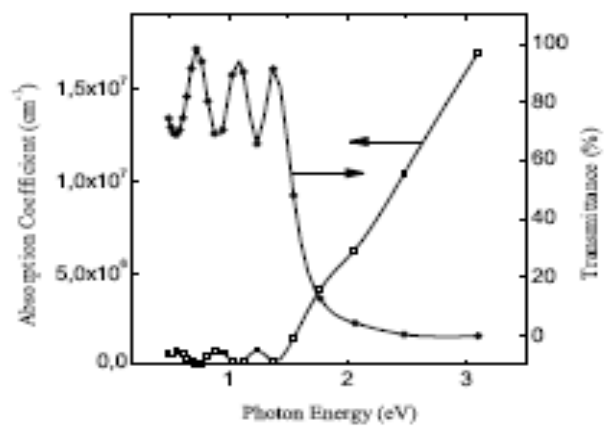

**Figure 2:** Transmittance and absorption coefficient spectrum of the CdTe thin films.

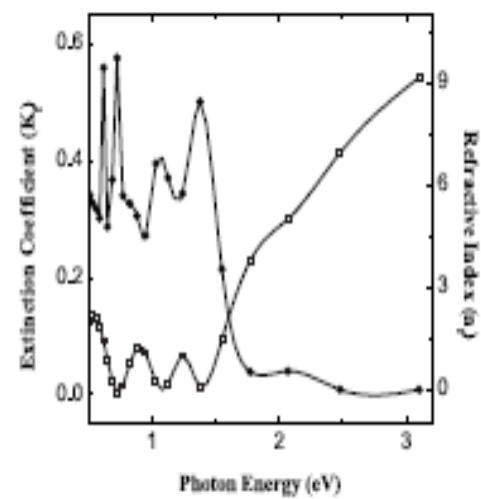

**Figure 3:** Dependence of $\alpha h\upsilon$ on photon energy for the CdTe thin films.

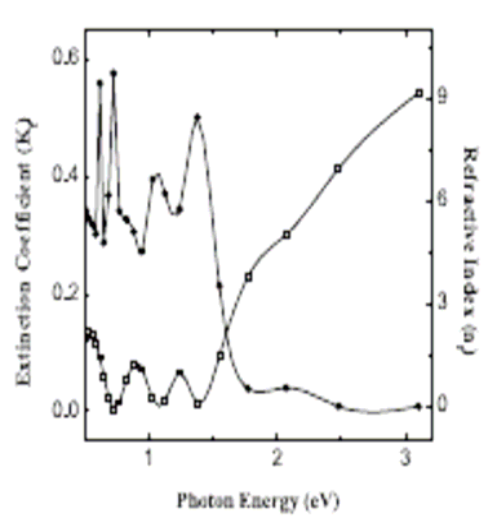

**Figure 4:** The variation of extinction coefficient and refractive index with wavelength for the CdTe thin film of thickness 5000Å.

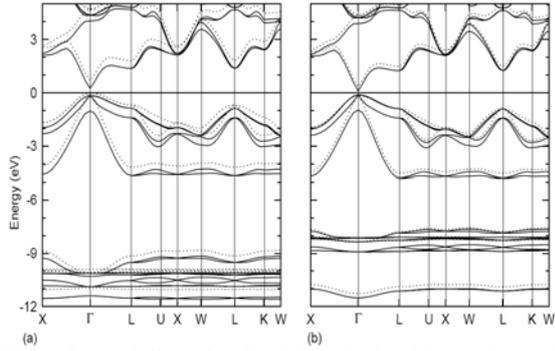

**Figure 5:** Band dispersion for CdTe calculated by (a) LDA+U and (b) LDA with (solid line) and without (dotted line) SO coupling.

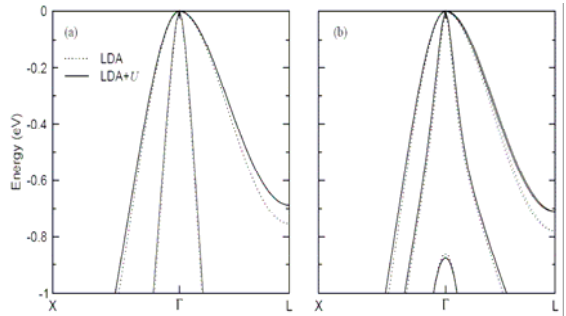

**Figure 6:** Band dispersion in the close vicinity of the $\Gamma$ point calculated from the LDA and LDA+U approaches (a) without and (b) with the SO coupling. The Fermi level is set at zero energy.

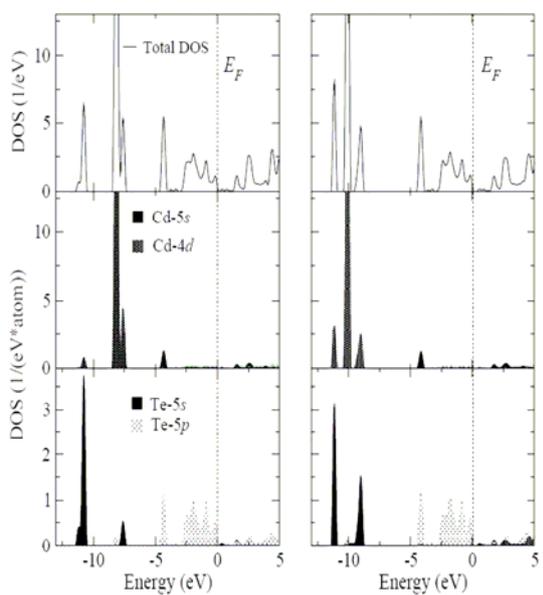

**Figure 7:** Total and partial DOS for Cd and Te atoms in CdTe films calculated from DFT within LDA (left panel) and LDA+$U$ (right panel). The Fermi energy ($E_F$) is set at zero of energy and is plotted by dotted lines.

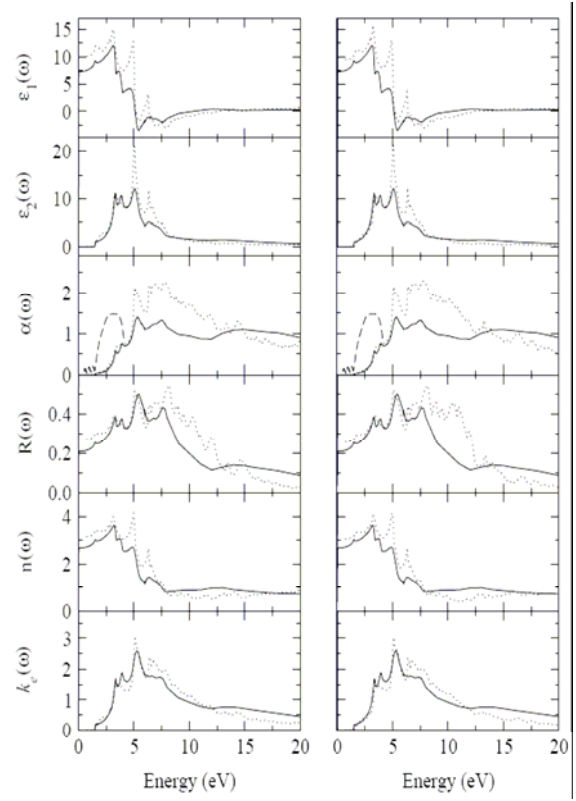

**Figure 8:** Optical spectra of CdTe thin films calculated (dotted lines) by VASP within the LDA (left column) and LDA+$U$ (right column) approximations. The absorption coefficients measured in this work (dashed line) and available from the other experiments [67] (solid lines) are given in [cm$^{-3}$] divided by $10^5$.

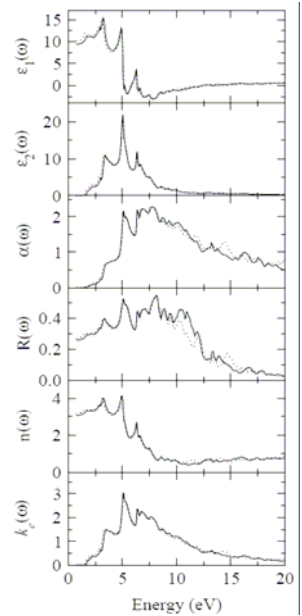

**Figure 9:** Comparison of the optical spectra of CdTe calculated within LDA+$U$ (solid lines) and LDA (dotted lines). $\alpha$ is given in [cm$^{-3}$] divided by $10^5$.

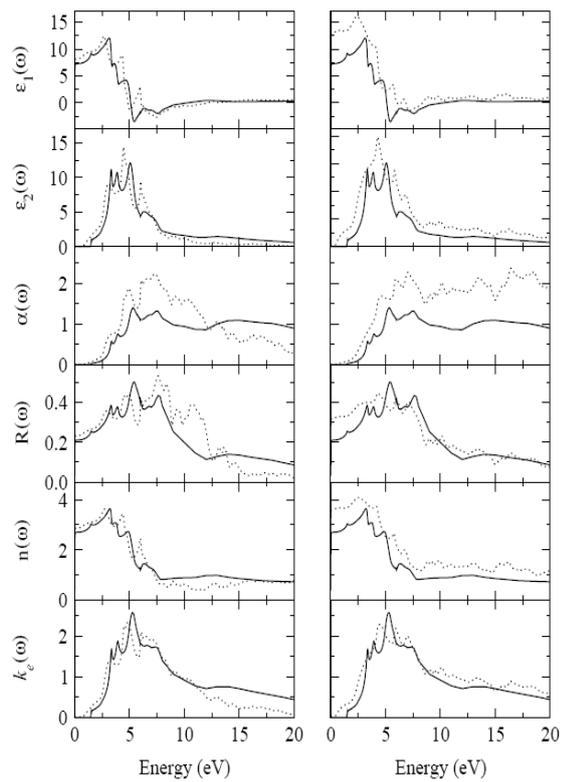

**Figure 10:** Optical spectra of CdTe calculated (dotted lines) by FP-LMTO method within the LDA+$U$ approximation without SO coupling (left column) and with SO coupling (right column). Solid line plots are the experimentally determined optical spectra in Ref. [67]. $\alpha$ is given in [cm$^{-3}$] divided by $10^5$.